\documentclass[pre,twocolumn,amsmath,amssymb,showpacs]{revtex4}
\def\dbar{{\mathchar'26\mkern-12mu d}}
\usepackage{graphicx}
\usepackage{dcolumn}
\usepackage{soul}
\usepackage{bm}

\begin{document}

%
\title{Dissipation and Entropy Production in Deterministic Heat Conduction of Quasi-one-dimensional Systems}

\author{Gary P. Morriss$^{~\star}$ and Daniel P. Truant}

\affiliation{School of Physics, University of New South Wales, Sydney NSW 2052, Australia}



\begin{abstract}
   We explore the consequences of a deterministic microscopic thermostat-reservoir contact mechanism. 
   With different temperature reservoirs at each end of a two-dimensional system, a heat current is produced and the system has an anomalous thermal conductivity. 
   The microscopic form for the local heat flux vector is derived and both the kinetic and potential contributions are calculated. 
   The total heat flux vector is shown to satisfy the continuity equation.
   The properties of this nonequilibrium steady state are studied as a function of system size and temperature gradient identifying key scaling relations for the local fluid properties and separating bulk and boundary effects.  
   The local entropy density calculated from the local equilibrium distribution is shown to be a very good approximation to the entropy density calculated directly from velocity distribution even for systems that are far from equilibrium.
   The dissipation and kinetic entropy production and flux are compared quantitatively and the differing mechanisms discussed within the BGY approximation.
   For equal temperature reservoirs the entropy production near the reservoir walls is shown to be proportional to the local phase space contraction calculated from the tangent space dynamics. 
   However, for unequal temperatures, the connection between local entropy production and local phase space contraction is more complicated.
\end{abstract}


\maketitle  

\section{Introduction}

   The study of heat conduction in low dimensional systems has concentrated on one-dimensional lattices \cite{PC00, GNY02, L03, FPU50} and simplified hamiltonian models that are amenable to solution \cite{EY04, CE09, CEM09}.
      The Hamiltonian models often consist of energy storage devices which couple to each other through the motion of tracer particles that carry the energy.
   Exceptions to this have been rare, but one of these is the study of particle based systems using hard disks by Deutsch and Narayan \cite{DN03}, and others \cite{Lip07}. 
   The other common feature of these model systems is that the thermal reservoirs are stochastic and are thus sources of particles with velocities chosen from some distribution and the reservoir is not mechanically coupled to the system.
   This element of randomness has limited the approaches that could be used to study heat conduction as a dynamical system \cite{Ott}.

   In 2007 a new deterministic thermal reservoir was introduced that coupled the quasi-one-dimensional system of hard disks to temperature reservoirs by changing the collision rule at the reservoir boundary \cite{TM07}. 
   For a collision with a reservoir boundary the tangential $y$-component of momentum is unchanged but the normal $x$-component after collision becomes
\begin{equation} \label{wallcoll}
p'_{x} = \epsilon  p_{res}-(1- \epsilon)  p_{x},
\end{equation}	 
where $p_{res}$ is a reservoir momentum related to the reservoir temperature by $p_{res}=\sqrt{ 2T_{res}}$ and $\epsilon$ is a reservoir coupling parameter.
   As $\epsilon \rightarrow 0$ the system decouples from the reservoir and the boundary becomes a hard wall, and as $\epsilon \rightarrow 1$ the incoming momentum is replaced by the reservoir momentum.
   A recent study of this system in contact with two reservoirs of the same temperature \cite{MT12} has shown that the active mechanical coupling leads to entropy production near each reservoir which then flows into the reservoir.
   These effects are local and involve a limited number of boundary layer particles regardless of the system size.

   Molecular dynamics simulations have proved a very effective means of testing theoretical approaches to the study of fluids both in equilibrium, and nonequilibrium steady states \cite {EM2}. 
   Given a particular atomic pair interaction, the results are free of approximations, with an accuracy limited only by statistical considerations. 
   It is usual to use the equipartition theorem to define the {\it kinetic temperature}, so in a system of $N$ particles in $d$ spatial dimensions the translational kinetic energy is $\tfrac {kT} {2}$ per degree of freedom.
   We define instantaneous local temperatures for each particle so that in two spatial dimensions $T_{i,x}=p_{i,x}^{2}/m$ and  $T_{i,y}=p_{i,y}^{2}/m$, and then the instantaneous system temperature is
\begin{equation} \label{kintemp}
T = \frac {1} {2N} \sum_{i=1}^{N} (T_{i,x}+T_{i,y})= \frac {1} {N} \sum_{i=1}^{N} \frac {\mathbf {p}_{i}^{2}} {2m},
\end{equation}	 
\noindent
   In the absence of a temperature gradient the average $\left< T \right>$ gives the system temperature, but when there is a temperature gradient the local time averages $\left<T_{i,x} \right>$ and $\left< T_{i,y} \right>$  give the local temperatures which will be used to determine the temperature profile inside the system.
    The difference between the local components of the local temperature can be used to give a measure of the deviation from local thermodynamic equilibrium. 
      We use $T_{i,x}$ and  $T_{i,y}$ to define a local {\it operational} temperature for a nonequilibrium system, but it is more usual to attempt to connect the average {\it local} kinetic temperature and the {\it local} thermodynamic  temperature.
   The relationship between this {\it kinetic} and {\it thermodynamic} temperature is only beginning to be explored  \cite{B00,E89,MR99,R05}.

   For a QOD system we can extended the idea of a kinetic temperature for a single atom to define the {\it local kinetic temperature} in the average volume element occupied by the particle, as the order of the particles is fixed. 
   Then we extend this idea to define other local thermodynamic properties using the properties of the particle that occupies the volume element. 
   In particular the velocity distribution of the particle can be used to define the velocity distribution of the volume element and then the thermodynamic properties that are derived from it.

   For equilibrium systems we expect a well-defined thermodynamic limit so that average properties converge with the inverse of the system size.
   However, for nonequilibrium systems a thermodynamic limit is not useful. 
   If the temperature of the two reservoirs are fixed and the system size increased then the temperature gradient goes to zero. 
   If the gradient is fixed then it is usual to observe a hydrodynamic instability at some threshold system size where the system behaviour abruptly changes. 
   Given this difficultly we consider sequences of {\it equivalent} nonequilibrium systems of different sizes or different gradients to obtain scaling relations which remain correct in the regime before the onset of hydrodynamic instability.
   

\section{The Model System}

   In this section we introduce the simple deterministic microscopic model for thermal coupling of a system of hard disks to a reservoir which can be used with either, an equilibrium system or nonequilibrium steady state, to supply or remove heat. 
   As this microscopic model couples mechanically and deterministically to the system it may be probed as would a standard dynamical system and energy and kinetic entropy flows calculated numerically without approximations.
   The mechanism has been introduced previously \cite{TM07}, and studied by computer simulation and kinetic theory \cite{MCA08,KM09}. 

\subsection{System Dynamics}

\begin{figure}[htb]
\begin{center}
\caption{Schematic presentation of an $N$ hard-disk quasi-one-dimensional (QOD) system. The height $L_y$ is sufficiently small that the disks cannot pass one another. We choose the coordinate origin to be located at the bottom left corner of the system, and the periodic upper and lower system boundaries at $y=0,L_y$ are denoted by dashed lines. The boundaries at $x =0$ and $x=L_x$ are the hard walls of the reservoirs so this is a (H,P) QOD system.} 	\label{fig:QDS}
	\includegraphics{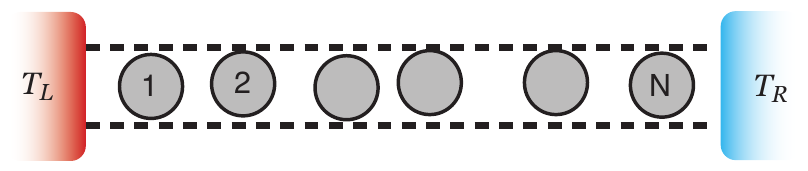}
\end{center}
	\end{figure}

   The quasi-one-dimensional (QOD) system introduced by \cite{TM03} can be modified to interact with an idealized {\it heat  reservoir} in a deterministic and reversible way, to study both heat conduction in low dimensional systems \cite{L03}, and the Lyapunov spectra and mode structure. 
   The deterministic reservoir allows the calculation of the usual dynamical systems properties as well the thermodynamic properties.
   This system contains hard disks of diameter $\sigma$ (which we set equal to 1) in a narrow channel that does not allow the disks to change their positional order, see figure (\ref{fig:QDS}).
   Therefore any property of particle $i$ can be associated with the same {\it local} property in the volume element $V_{i} = L_{y}(\left< x_{i+1}-x_{i-1} \right>)/2$, centred at the average position of particle $i$, $\left<x_{i}\right>$, thus for example, the local density is the inverse of the average volume occupied by the particle $\rho_{i} = 1/V_{i}$.

   The equations of motion connecting the QOD system to the two reservoirs, one on the left-hand (LH) side at $x =0$ and the other on the right-hand (RH) side at $x=L_{x}$ define the {\it thermal contact} so when a particle collides with a reservoir wall the normal component of the momentum of the particle $p_{x}$ is changed as given in Eq. (\ref{wallcoll}).
   For the LH reservoir $p_{res}=p_{L}$ is the fixed value of the reservoir momentum and for the RH reservoir  $p_{res}=-p_{R}$ (note that the reservoir momentum is always directed into the system).
   The reservoir coupling parameter $\epsilon$ represents the strength of the coupling of the reservoir to the system. 
   If $\epsilon = 0$ there is no interaction with the reservoir, and if $\epsilon = 1$ the incoming momentum is completely replaced by the reservoir momentum.
   Here we use an intermediate value of $\epsilon = 0.5$ which provides an effective mix of the incoming momentum with the reservoir momentum. 
   The system has a volume $V = L_{x} L_{y}$ with a fixed width of $L_{y}=1.15\sigma$, and then $L_{x}$ varies with the number of particles $N$ to give the desired density $\rho = N \sigma^{2} / (L_{x} L_{y})$.
   As the QOD system is narrow enough to prevent particles interchanging their positions $L_{y} < 2 \sigma$, the order of the particles remains fixed (both $\sigma$ and the mass $m$ are set to one).
   The temperature profile is determined from the average components of the kinetic temperature of each particle (the average of Eq. \ref{kintemp}).

   To produce a nonequilibrium steady state it is sufficient to have reservoirs of different temperature on each side of the QOD system.
   The energy entering the system from a boundary with reservoir momentum $p_{res}$ during a collision with a particle of incoming momentum $p_{x}$ is  given by
\begin{equation}
\Delta e_{I} = \frac {1} {2} ({p'}_{x}^{2} - p_{x}^{2}) = \frac {\epsilon} {2}  [ \epsilon p_{res}^{2} + 2 (\epsilon - 1) p_{res} p_{x} + (\epsilon - 2) p_{x}^{2}].
\end{equation}
   The time average of this quantity gives the flux of energy {\it into} the system so for a total system energy balance $\Delta e_{L}$ must be equal in magnitude but opposite in sign to $\Delta e_{R}$.
   The energy flux is controlled by both $p_{res}$ and the value of $\epsilon$, going to zero as $\epsilon \rightarrow 0$ and the reservoirs become disconnected from the system.

  We choose the temperatures of the two reservoirs independently, so the temperature of the left-hand side reservoir is $T_{L} > 2$ and varies, and the temperature of the  right-hand side reservoir is kept constant at $T_{R}=2$. 
   This produces a temperature gradient of $\nabla T=(T_{R}-T_{L})/L_{x}$ so heat will flow from the left-hand reservoir to the right-hand reservoir. 
   An overview of simulation state points is given in Table. (\ref{statepoints}).

\begin{table}[htdp]
\caption{The simulation state points. For each density $\rho$ there is a small and large temperature gradient considered. Generally, $T_{R}=2$ and groups of simulations are carried out with either $\nabla T$ fixed varying $N$ (or equivalently $\rho$) or $N$ fixed varying $\nabla T$. }

\begin{center}
\begin{tabular}{|c|cc|} \hline
$\rho$  & small $\nabla T$ & large $\nabla T$   \\  \hline
$0.03$ &        $-0.000431$ & $-0.01380$  \\
$0.8$   &         $-0.01150$   & $-0.3680$   \\
\hline
\end{tabular}
\end{center}
\label{statepoints}
\end{table}

\subsection{Tangent Space Dynamics}\label{TSdyn}

   The tangent space dynamics changes due to the changed collision rule at each boundary \cite{MT12r} so that for a particular reservoir we have
\begin{eqnarray}
   \delta x' &=& \left( \epsilon \frac {p_{res}} {p_{x}} + (\epsilon -1) \right) \delta x \\
   \delta p'_{x} &=& ( \epsilon -1) \delta p_{x}. 
\label{TS_BC} 
\end{eqnarray}
where  $p_{res}$ is the value of the reservoir momentum.  
   Notice that both $p_{L}/p_{x1}$ and $p_{R}/p_{xN}$ are negative so the sign of $\delta x'$ is opposite that of $\delta x$.

   The time evolution of the QOD system in tangent space is an infinite product of pairs of tangent matrices for a free-flight $F_{i}$ and a collision $C_{i}$. 
   The tangent space dynamics for an arbitrary tangent vector $\delta \Gamma (0)$ can be written as $\delta \Gamma (t) = M(t) \delta \Gamma (0)$ where $M(t)$ is a product of $F$ and $C$ matrices and the Lyapunov exponents are the logarithms of the eigenvalues of
\begin{equation}\label{Lambda}
\Lambda = \lim_{t\rightarrow \infty} [M(t)^{T} M(t)]^{1/2t}.
\end{equation}
   For any free-flight or particle-particle collision the tangent matrix has a determinant that is equal to one.
   Only wall-particle collision matrices have a determinant that depends upon $\epsilon$ and differs from one so the dynamics become dissipative.
   The dissipation from a single collision with a reservoir $d_I$ is given by the determinant of the wall collision matrix $C^W_{I}$ as
\begin{equation}\label{walldet}
d_I = \det C_{I}^{W} = (\epsilon - 1)\left( (\epsilon - 1) +  \epsilon \frac {p_{res}} {p_{x}} \right)
\end{equation}
where $p_{x}$ is the incoming momentum of the colliding particle.
   Note that the determinant approaches one as  $\epsilon \rightarrow 0$, or as the wall becomes purely reflective, and as $p_{res} / p_{x}$ is negative, both terms in Eq. (\ref{walldet}) are negative ($0 \le \epsilon <1$), so their product $d_{I} > 0$. 
   Combining equations (\ref{Lambda}) and (\ref{walldet}), the sum of the Lyapunov exponents for the system becomes
\begin{eqnarray}
\sum_{j=1}^{2dN} \lambda_{j} = \ln (\det (\Lambda) ) 
= \nu_{L} \left < d_{L} \right> + \nu_{R} \left<  d_{R} \right>
\end{eqnarray}
where $\nu_{I}$ is the collision frequency for reservoir $I$.
   As $\det C_{I}^{W}$ depends on $p_{x}$ for both reservoirs, the angular brackets represent an average over the incoming distribution of $p_{x}$ which in itself depends on the values of $p_{res}$.
   Further, we can calculate separately the components of dissipation associated with each reservoir, and this gives more local information than is available from the sum of Lyapunov exponents.
   
\subsection{Microscopic Heat flux vector}

   For a system of spherical particles the microscopic representation for the instantaneous local heat flux vector at position ${\bf r}$ and at time $t$ is given by \cite{EM2,McL89}
\begin{eqnarray}\label{J_Q}
&&{\bf J}_{Q} ({\bf r},t) = \sum_{i=1}^{N} U_{i} {\bf v}_{i} \delta ({\bf r} -{ \bf r}_{i}) \nonumber \\ 
&&- \tfrac {1}{2} \sum_{i,j}^{N} {\bf r}_{ij} {\bf F}_{ij}\cdot ({\bf v}_{i} + {\bf u}({ \bf r}_{i}) - {\bf u} ( {\bf r}))   \nonumber \\ 
&& \times  \int_{0}^{1} d \lambda \delta({\bf r} -{ \bf r}_{i} - \lambda {\bf r}_{ij})
\end{eqnarray}
where $U_{i} = \frac {1} {2} m({\bf v}_{i}- {\bf u}({\bf r}))^{2}$ is the internal energy of particle $i$, ${\bf u}({\bf r})$ is the local streaming velocity at ${\bf r}$ and ${\bf u}({\bf r}_{i})$ is the local streaming velocity at the position of particle $i$.  
   For this system the local streaming velocity is zero everywhere.  
   We define the vectors $\mathbf {r}_{ij} =  \mathbf {r}_{j} -  \mathbf {r}_{i}$ and $\mathbf {p}_{ij} =  \mathbf {p}_{j} -  \mathbf {p}_{i}$.
   For hard core particles the interaction force is an impulse ${\bf F}_{ij} = ({\bf \hat r}_{ij} \cdot {\bf v}_{ij}) {\bf \hat r}_{ij} \delta (t- t_{ij})$, where $t_{ij}$ is the time at which a collision occurs between particles $i$ and $j$, therefore Eq. (\ref{J_Q}) becomes
\begin{eqnarray}\label {jq1}
&&{\bf J}_{Q} ({\bf r},t) = \sum_{i=1}^{N} U_{i} {\bf v}_{i} \delta ({\bf r} -{ \bf r}_{i})  \nonumber \\
&& - \tfrac {1}{2} \sum_{i,j}^{N} {\bf \hat r}_{ij} ({\bf \hat r}_{ij} \cdot {\bf v}_{ij}) {\bf \hat r}_{ij}\cdot ({\bf v}_{i}+{\bf v}_{j})\delta (t- t_{ij})  \nonumber \\
&& \times \int_{0}^{1} d \lambda \delta({\bf r} -{ \bf r}_{i} - \lambda {\bf r}_{ij}).
\end{eqnarray}
   Notice that if ${\bf v}_{i} + { \bf v}_{j}=0$ there is no collisional energy transfer so the transfer of energy is correlated with the fluctuations of the pair momentum away from zero. 
   In the integral, the delta function moves along a line from the position of particle $i$, ${\bf r}_{i}$, to the position of particle $j$, ${\bf r}_{j}$, as $\lambda$ goes from $0$ to $1$, so a reasonable proposition is to assign half the potential contribution to each particle.   
   This is analogous with assigning half the potential energy of interaction to each particle.
   We can make this more solid by considering the one strip approximation to the integral, that is $\frac {1}{2} \Delta (\delta({\bf r} - {\bf r}_{i}) + \delta({\bf r} - {\bf r}_{j})) $ where the strip width is $\Delta = || {\bf \hat r}_{ij} || = 1$. 
   Then Eq. (\ref{jq1}) becomes
\begin{eqnarray}\label {jq2}
&&{\bf J}_{Q} ({\bf r},t) = \sum_{i=1}^{N} U_{i} {\bf v}_{i} \delta ({\bf r} -{ \bf r}_{i}) \nonumber \\
&&- \tfrac {1}{4} \sum_{i,j}^{N} {\bf \hat r}_{ij} ({\bf \hat r}_{ij} \cdot {\bf v}_{ij}) {\bf \hat r}_{ij}\cdot ({\bf v}_{i}+{\bf v}_{j})\delta (t- t_{ij})  \nonumber \\
&& \times ( \delta({\bf r} -{ \bf r}_{i})+ \delta({\bf r} -{ \bf r}_{j}) )
\end{eqnarray}
   In this form it is clear how the individual contributions are assigned to each particle.
   The kinetic contribution is at ${\bf r}_{i}$ while there are two potential contributions, one at ${\bf r}_{i}$ and the other at ${\bf r}_{j}$.
   
   We will be interested in the total heat flux ${\bf J}_{Q} (t)$ obtained as the volume integral of Eq. (\ref{jq2}), and the local heat current ${\bf J}_{Q} ({\bf r}_{i},t)$ obtained from the integral over the volume assigned to particle $i$, that is the integral over $V_{i}$.
   For the total heat flux all delta functions are contained within the integration region so
\begin{equation}\label {jqT}
{\bf J}_{Q} (t) V = \sum_{i=1}^{N} U_{i} {\bf v}_{i}  - \tfrac {1}{2} \sum_{i,j}^{N} {\bf \hat r}_{ij} ({\bf \hat r}_{ij} \cdot {\bf v}_{ij}) {\bf \hat r}_{ij}\cdot ({\bf v}_{i}+{\bf v}_{j})\delta (t- t_{ij}).  
\end{equation}
   For the QOD system the local heat flux has potential contributions from two sources, either from a collision of particles $i$ and $i+1$ or from a collision of particles $i-1$ and $i$.
   The result is
\begin{eqnarray}\label {jqL}
&&{\bf J}_{Q} ({\bf r}_{i},t) V_{i} = U_{i} {\bf v}_{i}   \nonumber \\
- && \tfrac {1}{4} \sum_{j \in \{i-1,i+1\}}^{N} {\bf \hat r}_{ij} ({\bf \hat r}_{ij} \cdot {\bf v}_{ij}) {\bf \hat r}_{ij}\cdot ({\bf v}_{i}+{\bf v}_{j})\delta (t- t_{ij}) 
\end{eqnarray}
   Clearly the time averages of these two heat currents are the physically important quantities as the heat current must satisfy a continuity equation.
   But the continuity equation states that for the QOD system, on average, the same amount of heat passes through any vertical line regardless of its position.
   Returning to the heat current density in Eq. (\ref{J_Q}), we can define the heat current at some arbitrary $x$ value.
   Instantaneously, there is only a kinetic contribution if a particle has its coordinate $x_{i} = x$, and there is only a potential contribution if two particles collide where for one $x_{i} < x$ and for the other $x_{j} > x$ so that the line of delta functions in Eq. (\ref{jq1}) has one at position $x$.
   The time average of this instantaneous quantity must satisfy the continuity equation.
   
   It is easy to see that summing the RHS of Eq. (\ref{jqL}) over all particles $i$ gives the RHS of Eq. (\ref{jqT}), therefore
\begin{equation}\label {jqLT}
 {\bf J}_{Q} ({\bf r},t) V  = \sum_{i=1}^{N} {\bf J}_{Q} ({\bf r}_{i},t) V_{i} 
\end{equation}
so that the total heat flux ${\bf J}_Q ({\bf r},t)$ is the weighted sum of local contributions with weights $V_{i}/V$.
   Taking the time average of both sides of equation (\ref{jqLT}) to obtain $\left < J_{Q} ({\bf r}) \right >_{t}$ and $\left < J_{Q} ({\bf r}_{i}) \right >_{t}$, which must by the continuity equation be equal, thus equation (\ref{jqLT}) reduces to a trivial equality.


\section{The Generalized Gibbs relation: Kinetic Theory}
\label{theory}
   The kinetic-theory basis for nonequilibrium thermodynamics of the heat transport model in QOD systems has been discussed recently \cite{KM09}.
   The basic ingredient is the kinetic contribution to the  Boltzmann entropy $S(t)$ which is defined, up to a constant, to be
\begin{eqnarray}\label{Bent}
S(t)&=&\int d{\bf r}~s({\bf r},t)   \nonumber \\
&=& - \int d{\bf r} \int d{\bf v}~ f({\bf r},{\bf v},t)\ln f({\bf r},{\bf v},t).
\end{eqnarray}
where $s({\bf r},t)$ is the entropy density at position ${\bf r}$ at time $t$. 
   We have set Boltzmann's constant to unity ($k_B = 1$). 
   In kinetic theory the time evolution of the distribution function $f({\bf r},{\bf v},t)$ can be obtained from the Boltzmann equation which, without external forces, takes the following form 
\begin{equation}\label{Beq}
\frac{\partial f}{\partial t}+ {\bf v}\cdot \frac{\partial f}{\partial {\bf r}}=J[f],
\end{equation}
where $J[f]$ is the collision integral. 
   The distribution function is normalized as
\begin{equation}\label{dist}
\int d{\bf v}f({\bf r},{\bf v},t)=n({\bf r},t),
\end{equation}
where $n({\bf r},t)$ is the local number density of the system.

   The local entropy-balance equation can then be derived theoretically by substituting the time-derivative of $f$ from the Boltzmann equation, Eq.~(\ref{Beq}), into the derivative of the Boltzmann entropy, Eq.~(\ref{Bent}). 
   The result is expressed as
\begin{equation}\label{entbal}
\frac {\partial s}{\partial t} + \nabla\cdot {\bf j}_s = \sigma,
\end{equation}
where $\sigma$ is the entropy production.
   In the above the entropy flux ${\bf j}_s$ is given exactly by
\begin{equation}\label{entfluxden}
{\bf j}_s({\bf r},t) = - \int d{\bf v}~ {\bf v} f\ln f,
\end{equation}
but the BGK approximation for the collision integral \cite{Zwanzig} is needed to obtain a computable form for the entropy-production per unit volume $\sigma$, given by
\begin{eqnarray}\label{entprod}
\sigma({\bf r},t) &=& - \int d{\bf v}~J[f]\ln f \nonumber \\
&=& \nu ({\bf r}) k \int d{\bf v} \{f({\bf v}) - f_{loc}({\bf v})\} \ln f({\bf v}).
\end{eqnarray}
   The term $\nu ({\bf r})$ is essentially a parameter that is related to the local collision frequency given by $\nu _{0} n({\bf r}) \sqrt{T({\bf r})}$  (where $\nu _{0}$  is a constant that depends on dimensionality). 
   According to the H-theorem \cite{H-theorem} the local entropy production $\sigma({\bf r},t)$ must be non-negative  everywhere in the system.
     All the local quantities associated with the entropy balance equation Eq. \ref{entbal} have been defined through the Boltzmann equation and their definitions are exact within the validity of the Boltzmann equation, and are not restricted to the linear regime. 

   We consider a steady system with a temperature gradient along the $x$ direction where the average total momentum is zero, to provide a concrete theoretical description of the well-adapted linear irreversible thermodynamics for our system.  
   In the Chapman-Enskog expansion solution to first order in the gradient (that is Navier-Stokes order \cite{Dorfman}) the distribution function is given by $f=f_{loc}(1+\Phi)$ where  $f_{loc}$ is the normalized  local-equilibrium distribution function. 
   Where there is the possibility of a difference between the local values of the $x$ and $y$ temperatures, $T_{x}$ and $T_{y}$, we need to use an appropriately modified  local-equilibrium distribution function of the form
\begin{equation}\label{loc}
f_{loc} (x,{\bf v}) = \frac {m n} {2 \pi} \left( \frac{1} {T_x T_y} \right)^{1/2}  \exp{\left[- \frac{m}{2} \left( \frac{v_x^2 } {T_x} + \frac{v_y^2 }{T_y}\right )\right]}.
\end{equation}
   Here $n$,  $T_x$ and $T_y$ are the local number density and $x$ and $y$ components of temperature which are all functions of position $x$.  

   The deviation from  local equilibrium is obtained at the first Sonine approximation explicitly as \cite{Santos3}
\begin{equation}\label{NS}
\Phi(x,{\bf v}) = - \frac{m}{2} \frac{\kappa} {p} \left( \frac{1}{2}m\beta v^2 -2\right) v_x\frac{d \ln T}{dx}.
\end{equation}
in which $p$ is  the hydrostatic pressure and $\kappa$ is the thermal conductivity.
   Note that the spatial dependence of solution, occurs only through the hydrodynamic fields, which is a well known characteristic of a \textit{normal} solution to the Boltzmann equation \cite{Kim2}, and it follows that any velocity moments will possess similar dependence. 
   Since the local energy balance in the steady state implies that $\nabla\cdot{\bf j}_Q(x)=0$, the heat flux must be uniform, in contrast to the entropy balance in which the entropy flux is nonuniform but satisfies $\nabla\cdot {\bf j}_s =\sigma$.

   The entropy density is obtained at the level of the local-equilibrium approximation
\begin{eqnarray}\label{Bentdenx}
s_{loc}(x) &\simeq&  -\int d{\bf v} f_{loc}\ln f_{loc} \nonumber \\
&=& n \left[1 - \ln \left(\frac{m n}{2\pi}\right) + \frac {1} {2} \ln ( T_{x} T_{y}) \right]
\end{eqnarray}
which is reminiscent of the equilibrium Sackur-Tetrode equation except that here the hydrodynamic fields are local. 
   The entropy flux is given to Navier-Stokes order \cite{Jou01} as
\begin{equation}\label{entfluxdenx*}
{\bf j}_s(x) \simeq  -k_B\int d{\bf v}~ {\bf v} \Phi\ln f_{loc}= -\kappa T^{-1}\frac{dT}{dx}\hat x.
\end{equation}
   Assuming Fourier's law, one can cast the expression for the entropy flux into the form, ${\bf j}_s(x) = {\bf j}_Q(x)/T$, and the entropy production in the the steady-state can be calculated from 
\begin{equation}\label{entprodx}
\sigma(x) = \nabla\cdot \left(\frac {{\bf j}_Q} {T} \right) \simeq\kappa \left| \frac{d\ln T}{dx}\right|^2,
\end{equation}
which is clearly positive.
   The relation between the entropy flux and the heat flux may also be viewed as a generalized version of the equilibrium Clausius relation $\dbar Q = TdS$ extended to steady states.
   Whether this remains an equality, or becomes an inequality, is central to extending thermodynamics to  nonequilibrium steady-states.


\section{Equilibrium}
   
\subsection{Dissipation and Entropy production}
   In a previous study \cite{MT12} it was shown that an energy balance is achieved for a QOD system with reservoirs of equal temperature, but that entropy was produced near each reservoir which then flows from the system to the reservoir.
   The observation that a system designed to be "in equilibrium"  is actually dissipative is somewhat surprising, but in a sense this effect is restricted to a small region near each boundary and the bulk of the system can be regarded as equilibrium. 
   The entropy production is calculated from the numerically generated momentum distributions in the BGY approximation \cite{Zwanzig} and is strictly a kinetic contribution ignoring possible configurational contributions.
   In sub-section \ref{TSdyn}, a dissipation calculated from the tangent space dynamics of the system which is not limited to the kinetic regime, but is rather the decrease in phase volume which leads to a negative sum of Lyapunov exponents.
   This dissipation is again a boundary effect as it arises from the tangent space dynamics of the collision between the reservoir and its neighboring particle. 
   All other collision events in the dynamics give no contribution to the dissipation.

   The two-dimensional momentum distributions for each particle are calculated numerically on a grid $M \times M$ with $M = 301$ typically, and a resolution of $0.05$.
   The entropy, entropy flux and entropy production are all calculated numerically by integrating this distribution using Eqs. (\ref{Bent}, \ref{entfluxden},  \ref{loc}) with an error of the order of the square of the resolution, that is $0.0025$.
   The entropy flux and production are defined through the momentum distributions for particles and are expected to be good representations at low density in the {\it kinetic regime} where $\rho < 0.1$, but would neglect other potential contributions at higher density.
   The expression for the kinetic entropy flux is formally exact whereas the kinetic entropy production is determined in the BGK approximation.

   Both the dissipation and entropy production in the system with equal reservoirs is intensive, so largely independent of the system size, and only involves properties of the particles that are in contact with the reservoirs.
   The dissipation involves the collisions between the closest particle and the wall while the entropy production depends on the distribution of velocities for particles near the wall.
   The numerical convergence of these two quantities is quite different; the dissipation converges quite quickly with random fluctuations, the entropy production converges with the smoothness of the distributions increasing slowly and monotonically at disks $1$ and $N$ until a steady value is reached. 
   Statistically, the distribution function needs to be sampled sufficiently to obtain a reliable value over the whole two-dimensional space $(v_x,v_y)$ and the convergence of the local equilibrium distribution also depends on the convergence to the local properties $n(x)$,  $T_x (x)$ and $T_y (x)$.
   For $N=40$ and $\rho=0.03$ the majority of the entropy production $60\%$ comes from particles $1$ and $N$ with $40\%$ from the other particles (principally $2$ and $N-1$).
   At higher density, $\rho=0.8$ the relationship is very similar.

\begin{figure}[htb]
\begin{center}
\caption{The entropy production of particles $1$ and $N$ plotted as a function of the dissipation at the left-hand reservoir for an equilibrium QOD system of $40$ hard-disks with $T_{L}=T_{R}=2$ and $\rho =0.03$. The curve is parametrized by the reservoir coupling parameter $\epsilon$ with values of $\epsilon = 0, 0.1, 0.3, 0.5, 0.6, 0.7, 0.8$ and $0.9$. At $\epsilon=0$ the system is isolated from the reservoir and both the dissipation and entropy production are zero. As the system is symmetric the same result follows for the right-hand reservoir. All results are obtained from averages over $1 \times 10^8$ collisions per particle and the results for $\sigma$ are lower bounds on the correct result.} 	\label{Sproddiss}
	\includegraphics{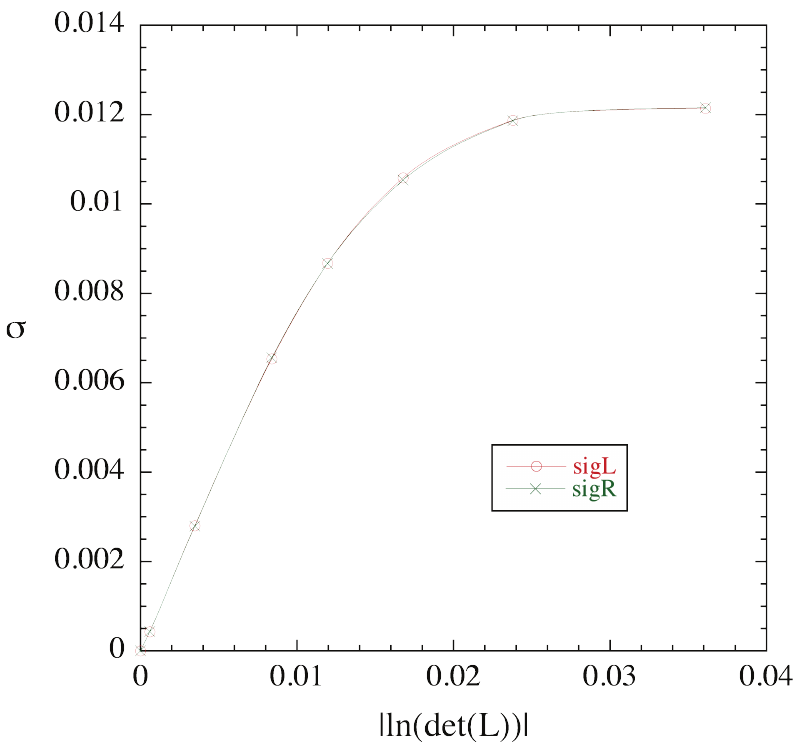}
\end{center}
\end{figure}

   As this system with equal reservoir temperatures shows both dissipation and entropy production and both properties appear to be intensive we look at both as a function of $\epsilon$, the strength of the coupling of the system to the reservoir. 
   From figure (\ref{Sproddiss}),  although these properties are of opposite signs, they appear to be proportional when parametrised by $\epsilon$.

\section{Nonequilibrium}

   We begin the study of nonequilibrium QOD systems by considering a system of $320$ hard disks at two different densities: a low density state of $\rho=0.03$ and a high density state of $\rho =0.8$.
   The temperature of the cold reservoir on the right-hand side is kept constant at  $T_{R}=2$, and different values of the temperature of the left-hand reservoir $T_L$ are used.
   The temperature gradient changes with the value of $T_{L}$ and the number of disks, so we consider both systems with constant temperature gradient and systems with constant $N$.
   Our purpose here is to identify approximate (or possibly exact) scaling relations for the properties of the system as functions of system size $N$ and temperature gradient $\nabla T$, with a view to separating bulk properties from surface properties (or boundary effects).

\subsection{Temperature profiles}

\begin{figure}[htb]
\begin{center}
\caption{(color online) The temperature profiles ($T_{x}$) for nonequilibrium QOD systems of $320$ hard-disks at densities of $\rho=0.03$ and $\rho =0.8$. In both cases the boundary temperatures are $T_{L}=130$ and $T_{R}=2$. The red symbols are the low density state $\rho=0.03$ and the blue symbols are the high density state $\rho=0.8$. Plotted as functions of particle number, the profiles appear almost indistinguishable, and thus imply that the profile is largely independent of density.} 	\label{T_x}
	\includegraphics{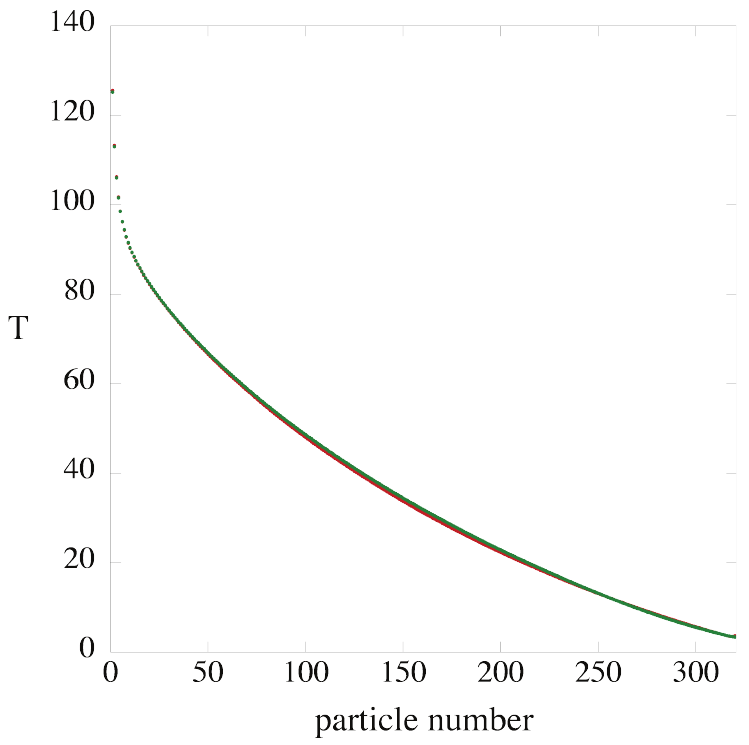}
\end{center}
\end{figure}

\begin{figure}[htb]
\begin{center}
\caption{(color online) The temperature difference profiles ($T_{x}(0.03) - T_{x}(0.8)$) for a nonequilibrium QOD systems of $320$ hard-disks with $T_{L}=130$ and $T_{R}=2$. Despite the seeming accuracy of Fig \ref{T_x} there are small systematic differences between the two temperature profiles. } 	\label{T_diff}
	\includegraphics{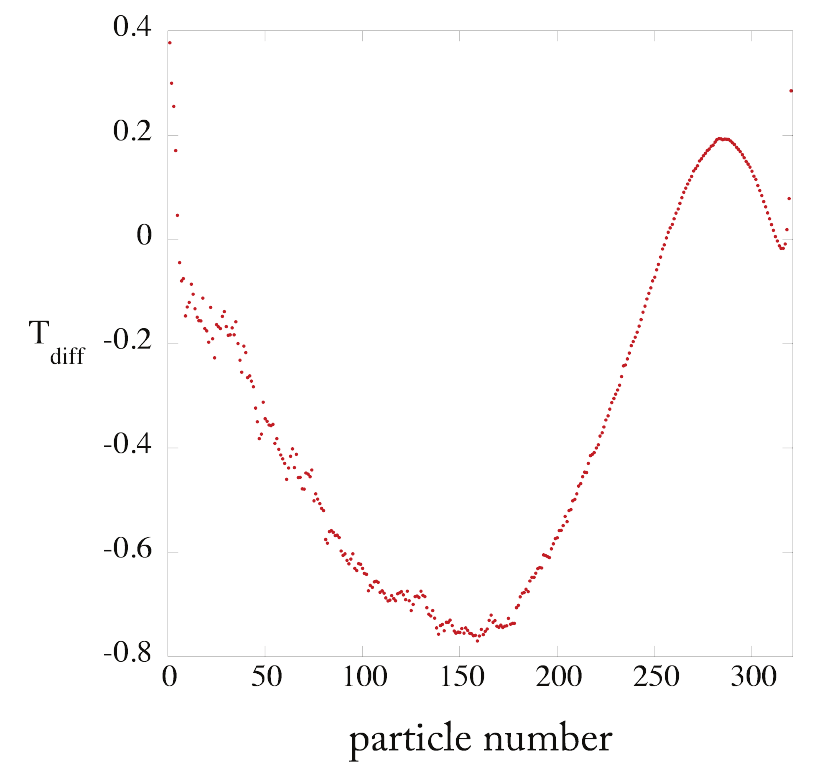}
\end{center}
\end{figure}

   We consider a nonequilibrium QOD system with $T_{L}=130 $ with $T_{R}=2$ at both the low density and high density states.
   Changing density, with fixed values of $T_L$ and $T_R$, changes the temperature gradient.
   The two temperature profiles for $T_x$ shown in Fig \ref{T_x} are surprisingly similar when plotted as functions of particle number, but we see in Fig. \ref{T_diff} that there are systematic small differences between these two profiles.
    
\begin{figure}[htb]
\begin{center}
\caption{(color online) The temperature profiles $T_{x}$ (red symbols) and $T_{y}$ (blue symbols) for $\rho =0.03 $ for a nonequilibrium QOD system of $320$ hard-disks with $T_{L}=130$ and $T_{R}=2$ plotted as a function of average particle position $\left< x\right >$. This profile looks quite different to the same system plotted as a function of particle number in Fig \ref{T_x} and appears to be approximately a cubic polynomial in $x$. } 	\label{T_03}
	\includegraphics{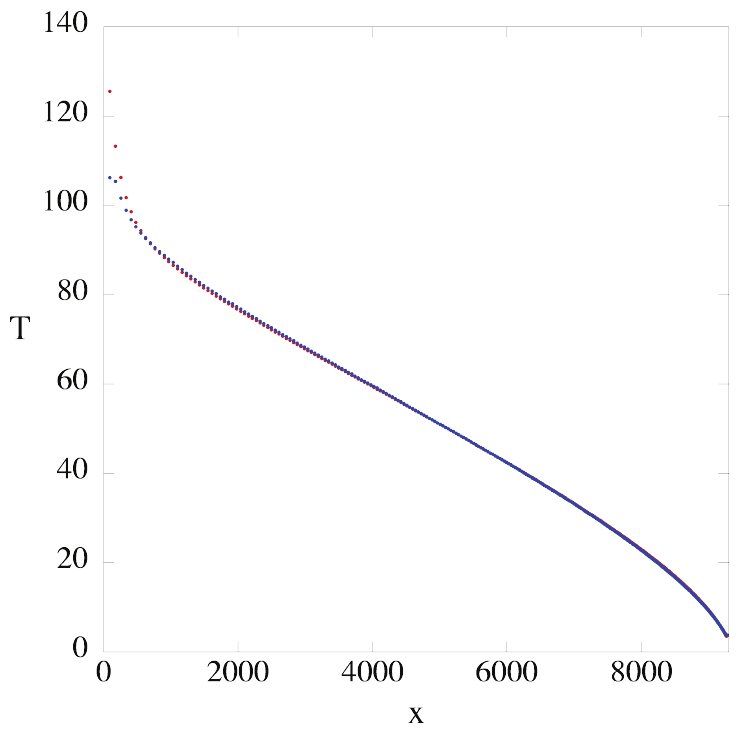}
\end{center}
\end{figure}

   Despite the small differences seen in Fig \ref{T_x}, these two profiles look quite different when plotted as functions of the average particle position $x$.
   For $\rho=0.8$ the density does not change greatly through the system so the profile as a function of particle number is very similar to the profile as a function of average particle position.
   However, for $\rho=0.03$ the profile changes significantly, becoming closer to a polynomial in $x$.
   Notice that in Fig. (\ref{T_03}) there is no place in the system where we could consider that the temperature profile is linear as expected from Fourier's law.

   Eckmann and Young \cite{EY04} derive an equation that the temperature profile should satisfy given by
\begin{equation}\label{stdtemp}
T^{\prime \prime}(x) T(x) = \gamma \left( T^{\prime}(x)\right)^{2}.
\end{equation}
   The parameter $\gamma = \alpha -1$ is equal to zero for a linear temperature profile.
   Here for all cases considered $T^{\prime \prime}(x)$ is so small that we cannot determine a non-zero value of $\gamma$ despite the obvious nonlinearity of the temperature profiles.
   
\begin{figure}[htb]
\begin{center}
\caption{(color online) The temperature difference profile $(T_{x}(x) -T_{y}(x))/T_{x}(x)$ as a function of scaled position $X = (i- \frac {1}{2})/N$ for nonequilibrium QOD systems of $80$, $160$, and $320$ hard-disks with the same temperature gradient $\nabla T=-0.01380$ at $\rho =0.03$.  The same graph for a density of $\rho=0.8$ is almost identical including its vertical scale.} 	\label{T_xy}
	\includegraphics{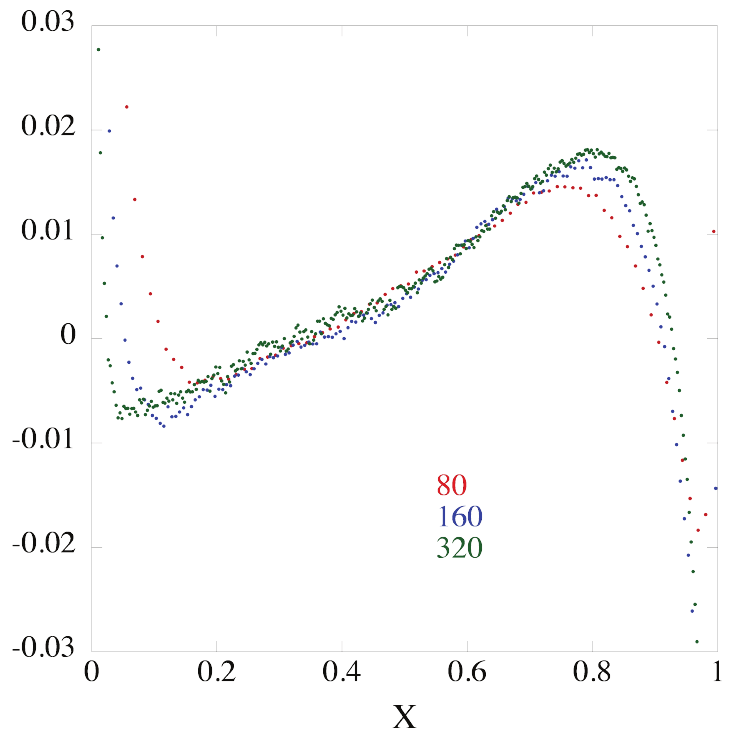}
\end{center}
\end{figure}

   The temperature difference profile $T_{x}(x) -T_{y}(x)$ to a good approximation satisfies a strong scaling relation that is independent of density and system size.
   To illustrate this we use systems of $80$, $160$ and $320$ disks at the same temperature gradient and a different position scaling on the horizontal axis. 
   We take the scaled particle position $X = (i- \frac {1}{2})/N$ where $i$ is the particle number on the horizontal axis, and plot $(T_{x}(x) -T_{y}(x))/T_{x}(x)$ on the vertical axis.
   In Fig. \ref{T_xy}  we notice that in the central part of the graph all curves coincide but at each end the curves change depending upon the system size.
   The figure for the high density case $\rho=0.8$ is almost identical, including the vertical scale.
   
\subsection{Energy balance}   

   The heat flux vector satisfies a continuity equation so must be, on average, constant everywhere in the system.
   It is composed of two parts: a kinetic part which dominates at low density, and a potential part which dominates at high density Eq. (\ref{jqL}).
   Regardless of the density, the total heat flux vector must match the energy that enters or leaves through the boundary (per unit length) by the collision mechanism Eq. (\ref{wallcoll}) as this is the only way energy can enter or leave the system.
   As before we can calculate the energy flux through the boundary in two different ways; first a direct calculation using the collision rule at the boundary $\Delta e_{L}/L_{y}$ or $\Delta e_{R}/L_{y}$ averaged over time, and second by calculating the average heat flux $J_{Q}$ which has both kinetic and potential components.
   In table (\ref{heatflux}) we present the components of the heat flux vector and show that the energy flow through the boundaries is consistent with the heat flux calculated using Eq. (\ref{jqL}).

\begin{table}[htdp]
\caption{The components of the heat flux vector $J_{Qx}$ for QOD systems of $N=320$ disks at two different densities and different left-hand temperatures $T_L$ with $T_{R}=2$. The superscripts $K$ and $\Phi$ signify the kinetic and potential components respectively. For the low density state $\rho = 0.03$ the heat flux vector is almost completely kinetic with a very small potential contribution. At the high density state $\rho = 0.8$ almost all of the heat flux vector is potential. Despite these differences the match between the total heat flux vector and the energy flow through the boundary is excellent. }

\begin{center}
\begin{tabular}{|cc|cccc|c|} \hline
$\rho$  & $T_{L}$ & $J_{Qx}^{K}$&  $J_{Qx}^{\Phi}$&  $J_{Qx}^{T}$ & $\Delta e_{L} / L_{y}$ \\  \hline
$0.03$ &       6  & $0.002917$ & $0.000098$  &  $0.003015$   &      $0.003015$   \\
$0.03$ &     10  & $0.006991$ & $0.000234$  &  $0.00723$     &      $0.007226$   \\
$0.03$ &     34  & $0.04604  $ & $0.001541$  &  $0.047581$   &      $0.047584$   \\
$0.03$ &     66  & $0.12404  $ & $0.004151$  &  $0.1282$        &      $0.1282$       \\
$0.03$ &    130 & $0.3398   $  & $0.01137$     &  $0.35117$     &      $0.35124$     \\  \hline
$0.8$   &        6  & $0.07574$   & $0.407122$  &  $0.48286$     &      $0.48426$     \\
$0.8$   &      10  & $0.1828 $    & $0.9719 $      &  $1.1547 $      &       $1.158 $        \\
$0.8$   &      34  & $1.2181$     & $6.2552$       &  $7.4733$       &      $7.4951$        \\
$0.8$   &      66  & $3.2850$    &  $16.547$       &  $19.832$       &      $19.889$        \\
$0.8$   &    130  & $8.9787$    &  $44.469$       &  $53.4477$    &      $53.6009$      \\
\hline
\end{tabular}
\end{center}
\label{heatflux}
\end{table}

\subsection{Local energy flux}

\begin{figure}[htb]
\begin{center}
\caption{(color online) The components of the local heat flux vector for a QOD system of $320$ disks at a density of $\rho =0.03$ with $T_{L}=130$ and $T_{R}=2$.  The red symbols are the kinetic contribution to the heat flux vector, the green symbols are the potential contribution and the brown symbols are the total heat flux vector which must be constant throughout the system. The red symbols are the local entropy flux times the local temperature which is equal to the kinetic contribution to the heat flux vector almost everywhere.} 	\label{JQ_flux_03}
	\includegraphics{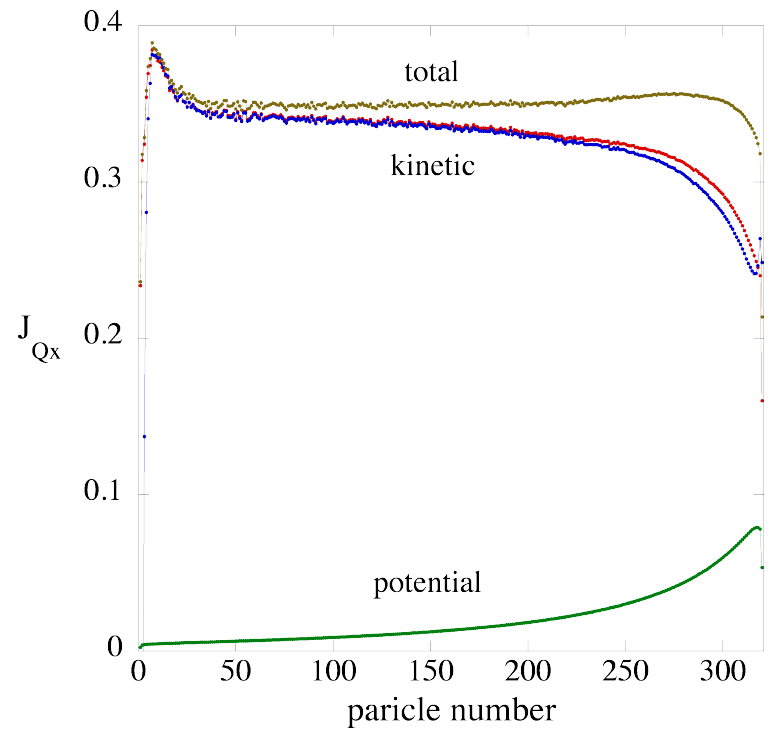}
\end{center}
\end{figure}

   In Fig. (\ref{JQ_flux_03}) we present the components of the heat flux vector for a QOD system of $320$ disks at low density where $T_{L}=130$ and $T_{L}=2$. 
   Despite the large temperature gradient at this low density state the major contribution to the heat flux comes from the kinetic term, but near the cold reservoir where the local density is greatest there is a contribution from the potential term that gives $25\%$ of the total.
   The total heat flux vector is close to constant throughout the system, except for a small peak near the hot reservoir and another small increase near the cold reservoir.
   Apart from these boundary effects, the continuity of the local heat flux vector is very good and changes in the kinetic contribution are compensated for by changes in the potential contribution.
   The local kinetic contribution is equal to the local temperature times the local entropy flux except in the region near the cold reservoir.
   If the local velocity distribution is exactly Gaussian at the local density and temperature then $J_Q (x) = T(x) j_s (x)$. 
   Here these two properties show systematic changes near the cold reservoir where the numerical velocity distribution must deviate sufficiently from the local equilibrium distribution.
   
\begin{figure}[htb]
\begin{center}
\caption{(color online) The components of the local heat flux vector for a QOD system of $320$ disks at a density of $\rho =0.8$ with $T_{L}=130$ and $T_{R}=2$.  The red symbols are the kinetic contribution to the heat flux vector, the green symbols are the potential contribution and the brown symbols are the total heat flux vector which must be constant throughout the system. The red symbols are the local entropy flux times the local temperature which is indistinguishable from the kinetic contribution to the heat flux vector everywhere.}   \label{JQ_flux_8}
	\includegraphics{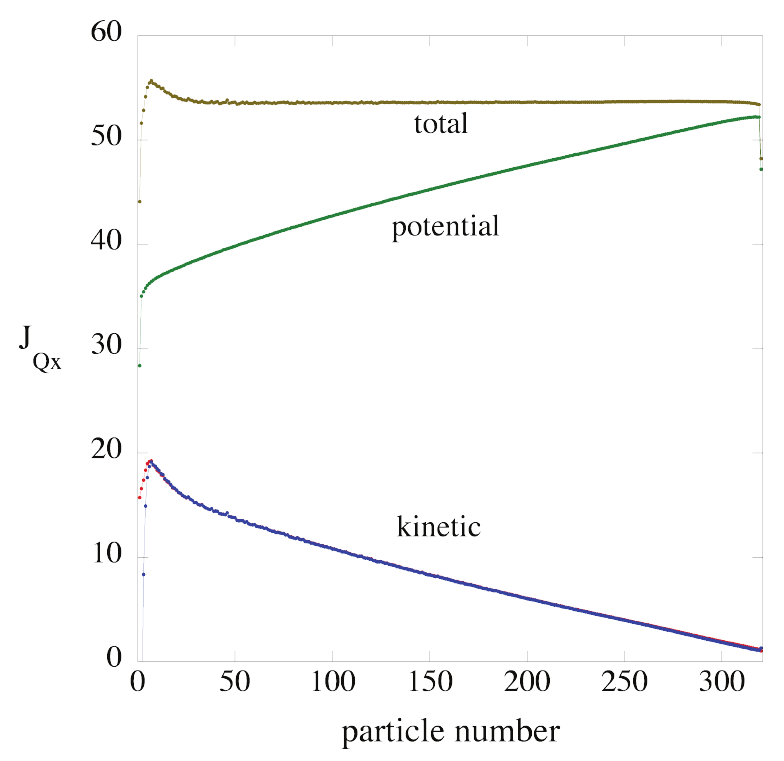}
\end{center}
\end{figure}

   For the higher density state the situation is somewhat different, see Fig. (\ref{JQ_flux_8}).
   Here the potential contribution to the local heat flux vector dominates the kinetic contribution and near the cold reservoir there is almost no kinetic contribution.
   Again the local total heat flux vector is constant throughout, apart from a small peak near the hot reservoir, and the local kinetic contribution to $J_Q (x)$ and $T(x) j_s (x)$ are indistinguishable everywhere.
   Indeed the agreement between the kinetic contributions is much better at high density than it is at low density in the kinetic region.
   
\subsection{Local collision frequency and entropy density}

\begin{figure}[htb]
\begin{center}
\caption{(color online) The theoretical BGK collision frequency $n(x) \sqrt{T(x)}$ plotted as a function of the directly calculated local collision frequency for systems of $80$, $160$ and $320$ disks at a density of $0.03$ and temperature gradient of $\nabla T = -0.0138$. In each case the externally applied temperature gradient is the same. The labels $H$ and $C$ signify the positions of the hot and cold reservoirs respectively.}	\label{fignu}
	\includegraphics{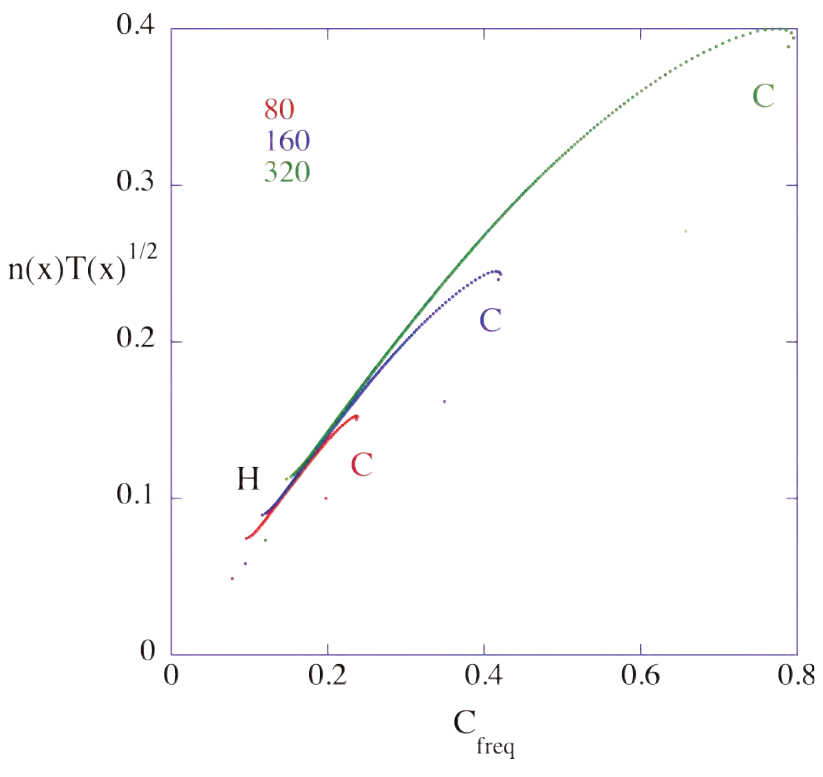}
\end{center}
\end{figure}

   In the kinetic regime we need the BGK approximation to obtain an estimate of the local entropy production and central to this is the local collisional relaxation frequency $\nu(x)$.
   We can test the proposition \cite{KD89} that the BGK local collision relaxation frequency is given by $\nu (x) = \nu_{0} n(x)  \sqrt{T(x)}$ in the kinetic regime by plotting it against the local collision frequency calculated directly.
   In Fig. \ref{fignu}, we present $n(x)  \sqrt{T(x)}$ as a function of the numerical collision frequency $C_{f}$ at a density of $0.03$ for three different system sizes ($N=80, 160, 320$) each with the same temperature gradient.
   From Fig. \ref{fignu} the resulting curve is the same and approximately linear, with deviations from linear  greatest near the cold reservoir where the density is highest and potential contributions are larger.
   Similarly, if we keep the system size fixed at $N=320$ and consider different values of $T_L$, and thus different $\nabla T$, the resulting curve is the same, and is the same as that shown in Fig. \ref{fignu}.
   The results are consistent with a single functional form $n(x)  \sqrt{T(x)}  = \frac {2} {3} C_{f}$.
   If we equate BGY relaxation frequency $\nu (x)$ with the numerical collision frequency $C_{f}$ then we have $\nu (x) =  \frac {3} {2}  n(x)  \sqrt{T(x)}  = C_{f}$.
   While we might expect that $n(x)  \sqrt{T(x)}$ is proportional to $C_{f}$, a universal linear relation is surprising. 
   At high densities this proportionality breaks down completely and these properties are not related in the same way.
   
  In section (\ref{theory}), the local entropy-balance equation was derived theoretically as Eq.~(\ref{entbal}), which equates the local change in entropy density to the divergence of the entropy flux and the entropy production $\sigma$.
  In a steady state the local entropy density does not change so $\nabla\cdot {\bf j}_s = \sigma$.
  The kinetic entropy flux is known exactly through Eq.~(\ref{entprod}) but the entropy production involves an unknown local relaxation frequency $\nu (x)$ that is the essential ingredient in the BGK approximation for the collision integral.
   Using the numerical results for the entropy flux and the entropy production (assuming that $\nu (x) = \nu_{0} n(x) \sqrt{T(x)}$) we can estimate $\nu_0$ as the slope of a best-fit line.
   The entropy balance equation can be wrtiten as
\begin{equation}\label{entprod}
\frac {\partial} {\partial x} j_{Sx}(x) = \nu_0 n(x) \sqrt{T(x)}  \int d{\bf v} \{f({\bf v}) - f_{loc}({\bf v})\} \ln f({\bf v}),
\end{equation}
where the left-hand side can be calculated, and all of the right-hand side, except for $\nu_0$, can be calculated, so the self consistent  $\nu _{0}$ value is the slope of the line.  
   We exclude the region nearest tho two reservoirs (approximately $10$ particles on each side), and the resulting curve is approximately linear with slopes given in Table. (\ref{nu_0}) for different values of the left-hand reservoir temperature $T_L$.
   This is a direct test of the accuracy of the BGY approximation, or a self-consistent calculation of $\nu_0$.
   The results are similar for all temperatures but different to the value obtained by relating $\nu (x)$ with the numerical collision frequency.

\begin{table}[htdp]
\caption{(color online) The self consistent estimate of the value of $\nu_0$ from the local entropy flux and BGY entropy production using QOD systems of $320$ disks at a density $\rho = 0.03$ with the right-hand  reservoir temperature of $T_{R}=2$ for a range of values of the left-hand reservoir temperature $T_{L}$.  }
\begin{center}
\begin{tabular}{|c|c|} \hline
$T_{L}$    &  $\nu_{0}$      \\  \hline
$18$         &  $ 0.295  $      \\
$34$         &  $ 0.232  $      \\
$66$         &  $ 0.229  $      \\
$130$       &  $ 0.218  $      \\
\hline
\end{tabular}
\end{center}
\label{nu_0}
\end{table}%

\begin{figure}[htb]
\begin{center}
\caption{(color online) The local entropy density $s(x)$ calculated from the numerical momentum distributions (blue symbols) and calculated from the local equilibrium distribution (red symbols) plotted as a function of the particle number for systems of  $320$ disks at a density of $0.03$. The label on each curve is $T_L$. For each temperature gradient the agreement is excellent. }	\label{s_den}
	\includegraphics{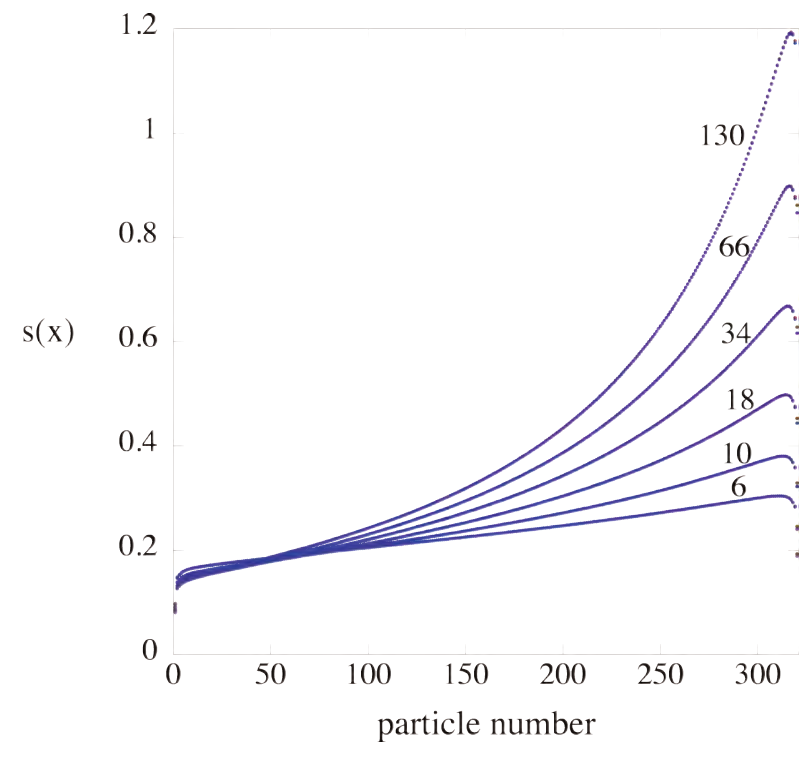}
\end{center}
\end{figure}

   While the momentum distributions for the particles cannot be exactly Gaussian as this nonequilibrium system supports an energy current, the deviations from Gaussian are at best only subtle and the local entropy density calculated from the numerical momentum distributions is almost indistinguishable from the local entropy density calculated from the local equilibrium distribution.
   The results in Fig. (\ref{s_den}) illustrate this for systems of $320$ disks for a range of different temperature gradients.
   
\subsection{Local entropy production and flux}

   The entropy production for a nonequilibrium steady state will be generated throughout the system wherever the particle momentum distribution differs from the local equilibrium distribution.
   The continuity of the heat flux vector throughout the system ensures that there are deviations from the local equilibrium momentum distribution at the position of each particle giving rise to local entropy production everywhere.
   This ubiquitous entropy production leads to an associated entropy flux towards the reservoirs, in addition to that observed at equilibrium, which will bias and add nonequilibrium effects to the equilibrium baseline.
   While the momentum distributions for the particles are perturbed from Gaussian because of the energy current, the deviations from Gaussian are small and at best subtle.
   The numerical convergence of the integrations required for these two properties are quite different. 
   The local entropy flux converges very quickly while the local entropy production, which is necessarily positive, is quite slow to converge and can have negative regions which take a very long time to decay to zero. 
   Indeed the dominant boundary entropy production terms for particles $1$ and $N$ seem to increase uniformly as they  converge. 
   In other regions, away from the reservoirs, the local entropy production decreases as it converges.

\begin{figure}[htb]
\begin{center}
\caption{(color online) The local kinetic entropy production divided by $\nu $ plotted as a function of particle number for QOD systems of $320$ disks at a density of $0.03$ with the right-hand reservoir at temperature $T_{R}=2$ for various values of temperature gradient (or left-hand reservoir temperature $T_{L}$.}	\label{Sprod}
	\includegraphics{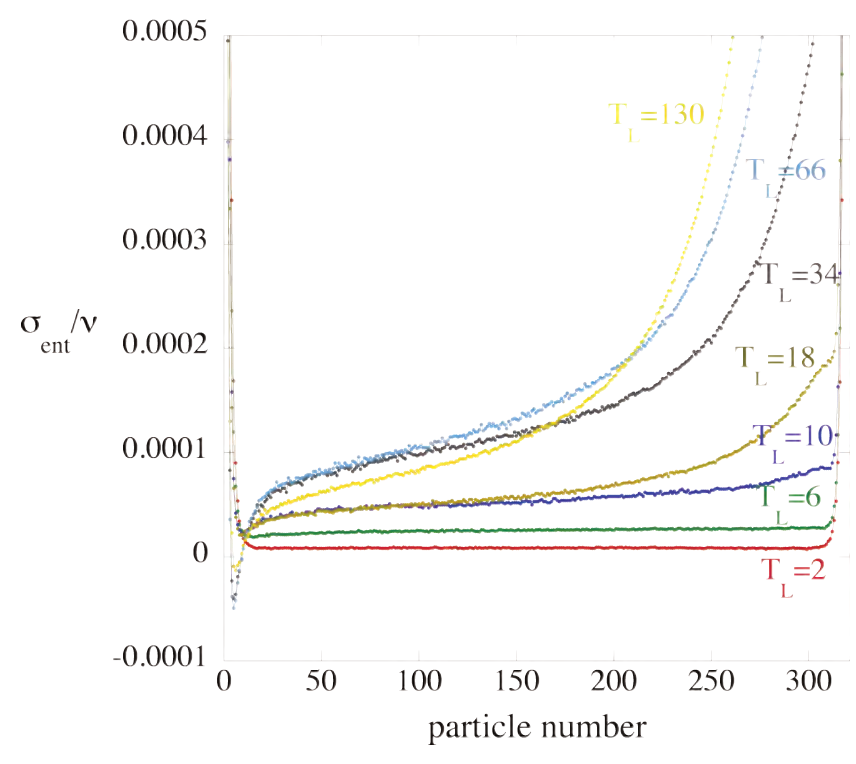}
\end{center}
\end{figure}

\begin{figure}[htb]
\begin{center}
\caption{(color online) The local kinetic entropy flux for the same systems as in Fig. (\ref{Sprod}). }	\label{Sflux}
	\includegraphics{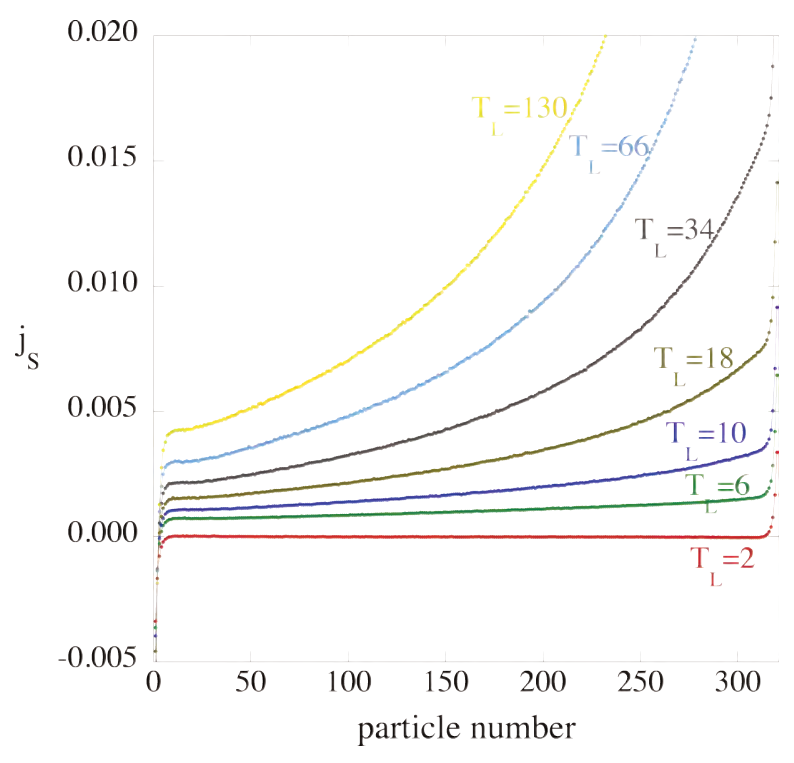}
\end{center}
\end{figure}

   For a QOD system of $320$ disks at density $\rho = 0.03$, shown in Fig. (\ref{Sprod}), we see systematic changes in the local entropy production $\sigma_{ent}/\nu$ as the value of $T_{L}$ changes. For $T_L$ up to values of about $10$ the changes are restricted to a lift in the baseline with a small positive slope.
   For larger values of $T_L$ the changes are less systematic and at large values $66$ and $130$ a negative region appears near the hot reservoir. 
      For the same system the local entropy flux shown in Fig. (\ref{Sflux}) has systematic changes for all values of $T_L$ up to the largest considered of $130$. 
   As before, the baseline is positive (away from the reservoirs) and increases with increasing particle number and with increasing $T_L$. 
   Clearly the fact that $j_S$ is positive implies the current of entropy is always directed to the right-hand cold reservoir in the centre and on the right-hand side but the value of the entropy flux for particle $1$ does not change with $T_{L}$ despite the fact that this is the boundary where the temperature changes.
   Similarly the entropy production at this wall is relatively constant with larger changes on the right-hand side.


   All calculations of entropy, entropy flux or entropy production consider only the kinetic contributions.
   These considerations are likely to be accurate for low density systems where we can assume that the potential contributions are small so we consider the low density QOD system with $T_{R}=2$ and a range left-hand reservoir temperatures $T_{L}$.    
   As the boundaries themselves, and the boundary layer, create entropy production and flux we ignore boundary effects by considering the bulk system to begin at $N=10$ and end at  $N=310$, for $320$ particle system. 
   We know that the heat flux vector is consistent with the heat flow across the boundaries so we work from the heat flux to calculate entropy fluxes and compare those with the ones calculated directly.
   
   The first row of Table (\ref{k102}) contains the numerical values of the local temperature and kinetic entropy flux $j_S$ calculated directly in the simulation.
   The local Clausius equality $j_{S} = J_Q /T$ suggests that the entropy flux can also be calculated from the ratio of the heat flux vector $J_Q$, and the local temperature, which are both accurately known in the simulation.
   If the Clausius equality holds then we expect that $j_{S,10}=J_{Q}/T_{10}$ and $j_{S,310}=J_{Q}/T_{310}$. 
   We can see that at particle $10$ near the hot reservoir this is a slight underestimate in $J_{Q}/T_{10}$ and at particle $310$ the value of $J_Q /T_{310}$ is an overestimate, possibly due to potential contributions as the density is higher near the cold reservoir.
   The results in Table (\ref{k102}) show that at low density, ignoring the boundary layer, the heat flux vector and the kinetic entropy flux agree reasonably well.
   For higher densities, such as $\rho = 0.8$ the potential contributions to the entropy flux that are implied by the values of $J_Q/T$ suggest that the kinetic entropy flux is only a small part of the total entropy flux and we would not expect agreement.
   
\begin{table}[htdp]
\caption{The local temperature, entropy and energy flux for a QOD system of $320$ disks at low density $0.03$ with reservoir temperatures $T_{R}=2$ and a range of values of $T_{L}$. We combine the reservoir and the nearest $9-10$ disks and consider the system explicitly at particle $10$ and particle $N=310$.}
\begin{center}
\begin{tabular}{|c|cc|c|cc|c|} \hline
$T_{L}$ & $T_{10}$ &   $j_{S,10}$   & $J_{Q}/T_{10}$&  $T_{310}$  &  $j_{S,310}$  &$J_{Q}/T_{310}$\\  \hline
$10$      & $7.4357$&   $0.001059$& $0.000971$      &  $2.0555$    &  $0.003394$ &  $0.003513$      \\
$18$      & $13.062$&   $0.00156$   & $0.001394$     &  $2.3201$    &  $0.007353$ &  $0.007849$       \\
$34$      & $24.215$&   $0.00214$   & $0.001968$     &  $2.7333$    &  $0.015299$ &  $0.017433$       \\
\hline
\end{tabular}
\end{center}
\label{k102}
\end{table}

\subsection{Phase space contraction and entropy production}

   The system is a nonequilibrium steady state with a local entropy production that we can estimate using the BGY approximation.
   The phase space contraction, as before for the equal reservoir case, can be calculated from the tangent space dynamics and will only contribute at the collisions of particles with the boundaries. 
   Here the different reservoir temperatures break the reflection symmetry and hence both the left and right-hand side phase space contractions will be different, as will the entropy productions for particle $1$ and $N$.
   We may imagine that the local entropy production and phase space contraction at each boundary may match separately but this is not what we observe. 
   The phase space contraction calculated from the tangent space dynamics of the particle nearest each reservoir, converges quite rapidly and although the dissipation will be an extensive quantity, the whole process takes places at the boundaries, mostly at the cold reservoir.
   At the cold reservoir the local density is highest and the collision frequency is highest but the average particle velocity is smallest.
   At the hot reservoir the local density is lowest and the collision frequency lowest but the average particle velocity is highest.

\begin{figure}[htb]
\begin{center}
\caption{(color online) Phase space contraction $\ln (\det (\Lambda))$ and entropy production $\sigma$ at disk $1$ (labelled L) and $N$ (labelled R) for QOD systems of $320$ disks with the right-hand reservoir temperature fixed at $T_{R}=2$ for a range of values of the left-hand reservoir temperature $2 \leq T_{L} \leq 130$. Panel (a) is the result for a density $\rho = 0.03$ and panel (b) is for a density $\rho = 0.8$. The filled symbols are values of the dissipation $\ln(\det(\Lambda))$ and the empty symbols are values of $\sigma$. The red symbols are for the left-hand reservoir and the green symbols are for the right-hand reservoir. }	\label{dSprod}
	\includegraphics{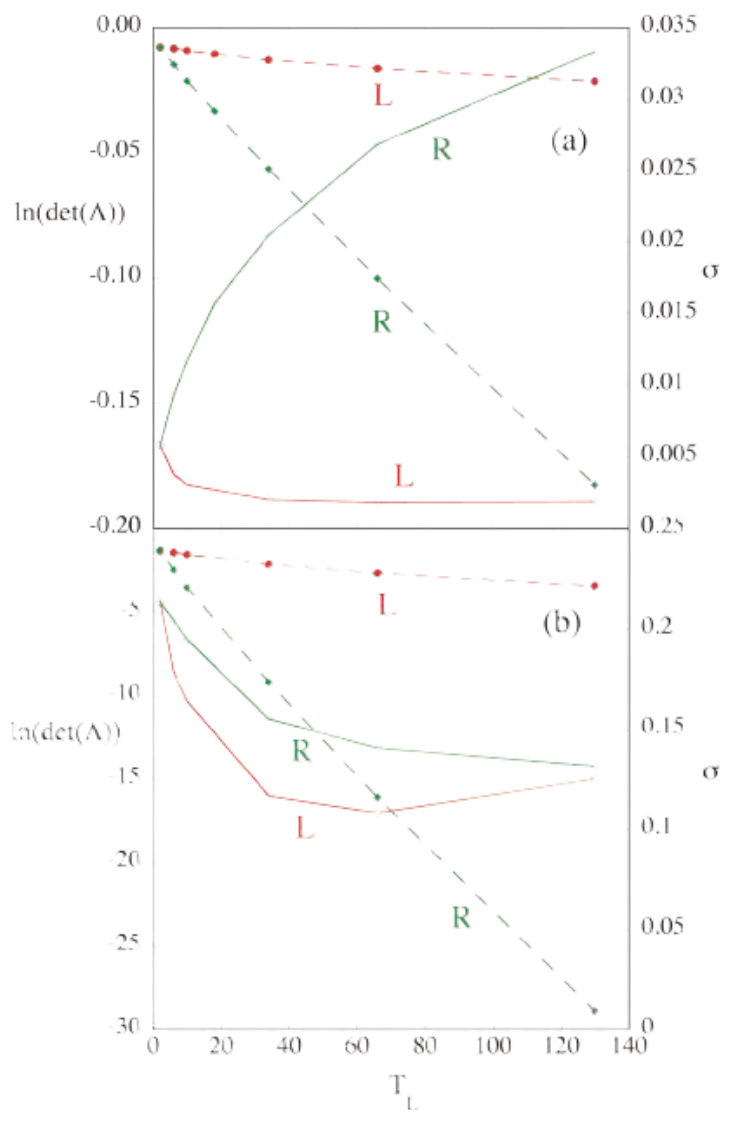}
\end{center}
\end{figure}

   Here we concentrate on a series of simulations at constant $N=320$ and $\rho = 0.03$ and $\rho=0.8$ to identify the dependence of the phase space contraction and boundary entropy productions $\sigma_1$ and $\sigma_N$ on the temperature gradient.
   The results for both densities are presented in Fig. (\ref{dSprod}) and the results are complicated.
   We note that space space contraction is negative whereas the entropy production is positive.

   Points to note about Fig. (\ref{dSprod}) are that when the reservoir temperatures are symmetric, with for example $T_{L}=T_{R}=2$ neither the phase space contraction nor the entropy production at $1$ or $N$ is zero, so this system is always dissipative. Both of these effects are caused by the boundary condition that is applied at each reservoir boundary.
   The phase space contractions for both densities are essentially linear in the reservoir temperatures $T_L$ and $T_R$, and the slope (or magnitude) is greatest at the cold reservoir where the collision rate is largest. The ratios of the slopes $T_R/T_L$ is also remarkably independent of density.

   The entropy productions at $1$ and $N$ are initially the same for $T_L=2$ but thereafter change systematically with $T_L$ but differently with density.
   At low density $\sigma_1$ decreases while $\sigma_N$ increases quickly with increasing $T_L$.
   Here it is tempting to associate $\ln(\det(\Lambda_R))$ and $\sigma_N$, but at the hot reservoir $\ln(\det(\Lambda_L))$ and $\sigma_1$ change in different directions.

   At high density both $\sigma_1$ and $\sigma_N$ decrease with increasing $T_L$.
   So the proportionality of $\ln(\det(\Lambda_R))$ and $\sigma_N$ observed for equal reservoir temperatures disappears when the temperatures differ, and is even more different at high density.

   The results presented in Fig. (\ref{dSprod}) are for the phase space contraction and entropy production as a function of $T_L$ or equivalently the temperature gradient.
   We have also studied these properties as a function of system size $N$ at fixed temperature gradient and here the graphs are very similar to those in Fig. (\ref{dSprod}).
   If we do a linear fit to the phase space contraction as a function of $N$ and find slopes $\alpha_L$ and $\alpha_R$ respectively (for L and R reservoirs), and then do a linear fit to the phase space contraction as a function of $T_L$ and find slopes $\beta_L$ and $\beta_R$ then we find the following relations in Table (\ref{slopes}).
   
\begin{table}[htdp]
\caption{The relationship between the slopes of linear fits to the phase space contraction as a function of $N$ and $T_L$.}
\begin{center}
\begin{tabular}{|c|c|c|c|c|} \hline
$\rho$ & $\alpha_{R}/\alpha_{L}$ & $\beta_{R}/\beta_{L}$ &    $\alpha_{L}/\beta_{L}$&  $\alpha_{R}/\beta_{R}$   \\  \hline
$0.03$    & $ 9.1$   &   $12.7 $     & $56$        &  $79 $         \\
$0.8$      & $10.4$  &   $12.95$   & $66.5$     &  $82.7$       \\
\hline
\end{tabular}
\end{center}
\label{slopes}
\end{table}

   The graphs of the entropy production $\sigma$ as functions of $N$ and $T_L$ are also very similar but as they are not linear we cannot repeat the analysis above.
   However, we can deduce from Table (\ref{slopes}) that at $\rho=0.8$ the functional form of $\ln(\det(\Lambda_L))$ as a function of $T_L$ can be obtained from the functional form as a function of $N$ by replacing $N$ by $66.5 T_L$.
   Similarly, the functional form of $\sigma_L(N)$ gives the functional form as a function of $T_L$ by replacing $N$ by $43 T_L$.
   These results are qualitatively the same for the low density $\rho=0.03$ with only small changes in numerical factors.
      
\subsection{Thermal conductivity}

\begin{figure}[htb]
\begin{center}
\caption{(color online) A log-log plot of the heat flux calculated in three independent ways as a function of $N$ for QOD systems at a density of $\rho=0.8$ and temperature gradient of $\nabla T = -0.368$ for a large range of system sizes. The right-hand reservoir temperature is fixed at  $T_{R}=2$ and $T_{L}$ varies such that $\nabla T$ remains fixed. The filled red circles are values of the heat flux into the system through the left-hand boundary $\Delta e_L/L_y$, the filled blue circles are the values of the heat flux out of the system through the right-hand boundary $\Delta e_R/L_y$, and the filled green circles are the values calculated from the heat flux vector $J_Q$ (\ref{jqL}). The line corresponds to $\Delta e = N^{1/2}$. }	\label{cond}
	\includegraphics{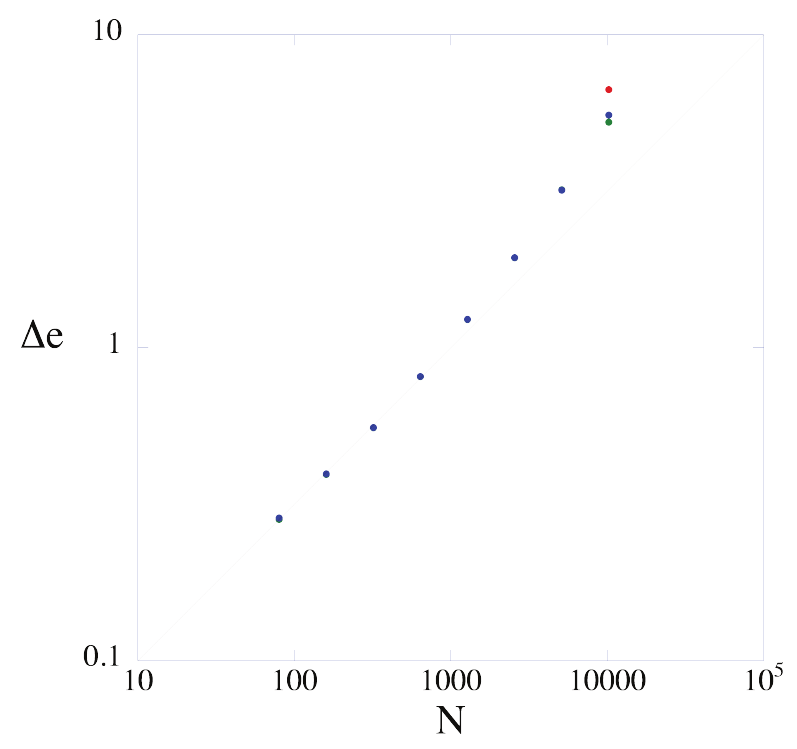}
\end{center}
\end{figure}

   Lower dimensional systems have long been shown to exhibit anomalous thermal conductivity with typically a power law divergence of the form $N^{\alpha}$ where $0.33 < \alpha \le \frac {1} {2}$.
   The QOD system is a restricted two-dimensional system where the average momentum is equal to zero, so we might expect two-dimensional behavior or perhaps marginally below two-dimensional.
   In Fig. (\ref{cond}) we have a result that suggests that $\alpha = \frac {1} {2}$ although extrapolation to large $N$ is always problematic.
   The numerical convergence of this system at fixed $N$ suggests that the initial value of the heat flux is high and then this converges slowly to a smaller value.
   Ideally, we may try to have the number of collisions per particle fixed for each system size but this is difficult for larger systems as the total run time increases at least as fast as $N^{2}$.
   Therefore the accuracy of the results decreases quickly for large system sizes.
   The consistency of the different routes to the heat flux gives a good indication of the convergence of the result, so the result for $N=10240$ is the least reliable of those reported in Fig. (\ref{cond}).

\section{Conclusion}

   We have investigated the consequences of a particular model microscopic coupling of a system to a reservoir. 
   The advantages of the model include its relative simplicity and the fact that it is deterministic which allows straightforward application of dynamical systems techniques. 
   Clearly the fact that the particles remain ordered from left to right, and this allows us to connect the properties of a volume element with the properties of a single particle, is a virtue of the quasi-one-dimensional system.
   In more realistic systems this is not possible as particles enter and leave Eulerian volume elements, but the generalisation is a technical rather than conceptual difficulty.

   The microscopic expression for the local heat flux vector derived here has been shown to satisfy the expected continuity equation for heat flow at both low density, where kinetic contributions dominate, and at high density where potential contributions dominate.
   The heat flux vector also agrees very well with the amount of energy entering the system from the left-hand reservoir and the amount of energy leaving via the right-hand reservoir.
   
   Strong scaling relations were obtained for the temperature profile as a function of particle number and for local temperature differences.
   These results suggest that the system behaves more simply than expected with regard to changes in density and temperature gradient.
   
   The entropy terms calculated here are all kinetic contributions and would not be expected to match at high densities although the kinetic contributions to the heat flux vector match the temperature times the kinetic contribution to the entropy flux.
   As potential contributions to the heat flux vector are large at high densities we would expect large potential contributions to the various entropy terms, including its production and flux, to be present.


\bibliographystyle{entropy}
\makeatletter
\renewcommand\@biblabel[1]{#1. }
\makeatother
\bibliography{}

\end{document}